\documentclass[
aps,
prc,
showpacs,
twocolumn,
superscriptaddress,
nofootinbib,
floatfix]
{revtex4-1}
\usepackage{graphicx,
longtable,
slashed,
hyperref,
bm,url,
braket,
xcolor,
xspace
}

\usepackage{comment}
\def\be{\begin{equation}}
\def\ee{\end{equation}}
\newcommand{\dg}{^{\dagger}}
\def\bee{\begin{eqnarray}}
\def\eee{\end{eqnarray}}

\newcommand{\lf}{\left(}
\newcommand{\rh}{\right)}

\begin{document}

\title{Structure of the two-neutrino double-$\beta$ decay matrix elements within perturbation theory}
%
%
\author{Du\v{s}an \v{S}tef\'anik}
\affiliation{ Comenius University, Mlynsk\'a dolina F1, SK--842 48, Slovakia}
\author{Fedor \v{S}imkovic}
\affiliation{ Comenius University, Mlynsk\'a dolina F1, SK--842 48, Slovakia}
\affiliation{ BLTP, JINR, 141980 Dubna, Moscow region, Russia}
\affiliation{ IEAP CTU, 128--00 Prague, Czech Republic}
\author{Amand Faessler}
\affiliation{ Institute of Theoretical Physics, University of Tuebingen,72076 Tuebingen, Germany}

\begin{abstract}
The two-neutrino double-$\beta$ Gamow-Teller and Fermi transitions are studied within
an exactly solvable model, which allows a violation of both spin-isospin SU(4) and isospin
SU(2) symmetries, and is expressed with generators of the SO(8) group.
It is found that this model reproduces the main features of realistic
calculation within the quasiparticle random-phase approximation with isospin symmetry restoration
concerning the dependence of the two-neutrino double-$\beta$ decay matrix elements on isovector
and isoscalar particle-particle interactions.  By using perturbation theory
an explicit dependence of the two-neutrino double-$\beta$ decay matrix elements on the
like-nucleon pairing, particle-particle $T=0$ and $T=1$, and particle-hole proton-neutron
interactions is obtained. It is found that double-$\beta$ decay matrix elements do
not depend on the mean field part of Hamiltonian and that they are governed by a weak violation of both
SU(2) and SU(4) symmetries by the  particle-particle interaction of Hamiltonian.
It is pointed out that there is a dominance of two-neutrino double-$\beta$ decay
transition through a single state of intermediate nucleus.
The energy position of this state relative to energies of initial and final ground
states is given by a combination of strengths of residual interactions.
Further, energy-weighted Fermi and Gamow-Teller sum rules connecting $\Delta Z = 2$ nuclei are discussed.
It is proposed that these sum rules can be used to study the residual interactions of
the nuclear Hamiltonian, which are relevant for charge-changing nuclear transitions.
\end{abstract}
\medskip

\pacs{
}
\maketitle

%
\section{Introduction}
%

With increasing sensitivity of double-$\beta$ decay ($\beta\beta$) experiments looking for a signal
of Majorana neutrino mass the problem of reliable calculation of neutrinoless double-beta decay
($0\nu\beta\beta$-decay) matrix elements $M^{0\nu}$ becomes more urgent \cite{verg12}. As far as is known
their value can not be related to any observable and must be calculated by using tools
of nuclear structure theory. Many sophisticated nuclear structure approaches including the large
basis interacting shell model \cite{poves12,horoi}, the interacting boson model \cite{barea13},
the projected Hartree-Fock-Bogoliubov method \cite{phfb}, the energy density functional method \cite{edf}
and various versions of the quasiparticle random phase approximation \cite{src09,newpar13,fae09}
were used to calculate them. The difference among obtained results are at the level
of factor 2-3 for particular nuclear systems \cite{verg12}. They can be attributed to truncation of the
nuclear Hamiltonian, many-body approximations, and various sizes of the single-particle model space.

The importance of the improvement of the calculation of the $0\nu\beta\beta$-decay nuclear matrix
elements is accepted worldwide. The quality of nuclear structure models can be improved
by complementary experimental information from related processes like two-neutrino double-$\beta$
decay ($2\nu\beta\beta$ decay), charge- and double-charge-exchange reactions, particle transfer
reactions, muon capture, etc.

The $2\nu\beta\beta$ decay \cite{verg12,hax84,doi85},
\begin{equation}
(A,Z) \rightarrow (A,Z+2) + 2 e^- + 2 {\overline{\nu}}_e,
\end{equation}
is a process fully consistent with the standard model of electroweak
interaction. So far it has been observed in twelve even-even nuclides
in which single-$\beta$ decay is energetically forbidden or strongly
suppressed \cite{barabash}. The measurement of $2\nu\beta\beta$-decay rates gives us
information about the product of fourth power of axial-vector coupling
constant $g_A$ and the  squared $2\nu\beta\beta$-decay matrix element
$|M^{2\nu}|$, which is a superposition of double Gamow-Teller (GT) and
double Fermi (F) matrix elements,
\begin{equation}
M^{2\nu} = M^{2\nu}_{GT} - \left(\frac{g_V}{g_A}\right)^2 M^{2\nu}_F.
\end{equation}
Here, $g_V$ is the vector coupling constant.

The observed values of $M^{2\nu}$ are used to study the nuclear structure
and nuclear interactions associated with the $0\nu\beta\beta$ decays.
The calculation of $M^{2\nu}$ requires a construction  of wave functions of
the even-even initial and final nuclei and of a complete set of $J^+=0^+$, $1^+$
states in intermediate odd-odd nucleus within a nuclear model. These wave
functions enter also in the evaluation of the neutrinoless double-$\beta$ decay matrix elements, which has a different form. 
The problem of a
reliable calculation of $M^{2\nu}$ is still not solved. Essentially,
calculations performed for nuclei of experimental interest
overestimate the $2\nu\beta\beta$-decay rate \cite{poves12,barea13}.
The shell model, which describes qualitatively well energy spectra, does
reproduce experimental values of $M^{2\nu}$ only by consideration of
significant quenching of the GT operator, typically by 60 to 70\%
\cite{poves12}.

In most quasiparticle random phase approximation (QRPA) calculations of
$M^{0\nu}$ the particle-particle interaction is adjusted so that
the $2\nu\beta\beta$-decay half-life is correctly reproduced
\cite{src09,newpar13}. As a result $M^{0\nu}$ values become
essentially independent of the differences in model space, nucleon-nucleon
interaction, and refinements of the QRPA method. Recently, a partial restoration
of isospin symmetry was achieved within the QRPA \cite{newpar13,fae09} by  separating the
particle-particle neutron-proton interaction into its isovector and isoscalar
parts, and renormalizing them each separately.  The isoscalar strength
parameter $g^{T=0}_{pp}$ is fit as before to $2\nu\beta\beta$-decay rates unlike
the isovector parameter $g^{T=1}_{pp}$, which is determined by the
requirement that $M^{2\nu}_F=0$ dictated by the isospin symmetry of the
nucleon-nucleon force. As a consequence, essentially no new parameter
is introduced as the strength of isovector particle-particle force is close to the
pairing force.

The Fermi and GT operators are generators of isospin SU(2) and spin-isospin SU(4)
multiplet symmetries, respectively. In the case of both symmetries being exact
in nuclei, the $2\nu\beta\beta$ decay would be forbidden as ground states
of initial and final nuclei would belong to different multiplets.
The isospin is known to be a good approximation in nuclei. Thus, it is
assumed that double Fermi matrix element is negligibly small and the
main contribution is given  by the double GT matrix element.
In heavy nuclei the SU(4) symmetry is strongly broken by the spin-orbit
splitting. But values of $M^{2\nu}_{GT}$ deduced from the observed
$2\nu\beta\beta$-decay rates are especially small for nuclei with large A.
It is worth noting that the $2\nu\beta\beta$-decay  transition to
ground state of final nucleus exhausts only about $10^{-4}$ of the double
GT sum rule \cite{voger88}. The existence of an (approximate) underlying
symmetry responsible for the suppression of the $2\nu\beta\beta$ decay,
which is assumed to be the SU(4) symmetry, justifies approaches
based on the perturbative breaking of this symmetry for construction
of wave functions of nuclear states participating in double-$\beta$ decay
transitions. To this category of methods belong the phenomenological
approach of Ref. \cite{desplan90} and various versions of the
proton-neutron QRPA.

Whether a discussed behavior of $M^{2\nu}_{GT}$ is a special property
of nuclei or just an artifact of the QRPA was discussed within
a schematic model which can be solved exactly and contains most
of the qualitative features of a realistic description \cite{vogel86}.
The vanishing of $M^{2\nu}_{GT}$ was identified with a dynamical
SU(4) symmetry of Hamiltonian. Later this model was exploited to
examine isovector and isoscalar proton-neutron correlations in the
case of GT strength and double-$\beta$ decay \cite{engel}.
In this paper we extend this schematic model to allow a violation
of both spin-isospin SU(4) and isospin SU(2) symmetries. The
main issue is to discuss explicit dependence of both $M^{2\nu}_{GT}$ and
$M^{2\nu}_F$ on the mean field and different components of residual
interaction by taking advantage of perturbation theory. We note
that a similar study, which has been found to be very instructive,
was performed for $M^{2\nu}_F$ by discussing violation of isospin symmetry
of Hamiltonian expressed with generators of the SO(5) group \cite{dusan13}.

%
\section{$2\nu\beta\beta$-decay rate and the importance of the energy denominators}
%

The $2\nu\beta\beta$-decay occurs as a second-order perturbation of the weak
interaction within the minimum standard model independently of whether neutrinos
are Dirac or Majorana. The effect of neutrino mixing and masses can be safely
neglected. The most favorable is the two-nucleon mechanism where the successive $\beta$ decays
of two neutrons in the even-even nucleus trigger the $2\nu\beta\beta$ decay.

The inverse half-life  of the $2\nu\beta\beta$-decay transition to the $0^+$
ground state of the final nucleus is given as follows:
\bee
\left[T^{2\nu \beta \beta}_{1/2}(0^+)\right]^{-1} =
\frac{m_e}{8 \pi^7 \ln{2}}(G_{\beta}m_e^2)^4I^{2\nu}\lf 0^+ \rh,
\eee
where $G_\beta = G_F \cos{\theta_C}$ ($G_F$ is Fermi constant and $\theta_C$ is the Cabbibo angle), $m_e$ is the mass of
electron, and
\bee
I^{2\nu}\lf 0^+ \rh &=&\frac{1}{m_e^9}\int_{m_e}^{E_i-E_f-m_e}F_0(Z_f,E_{e_1}) p_{e_1} E_{e_1} dE_{e_1} \nonumber \\
&\times&\int_{m_e}^{E_i-E_f-E_{e_1}} F_0(Z_f,E_{e_2}) p_{e_2} E_{e_2} dE_{e_2} \nonumber\\
&\times&\int_{0}^{E_i-E_f-E_{e_1}-E_{e_2}} E_{\nu_1}^2 E_{\nu_2}^2 {\cal A}^{2\nu} dE_{\nu_1}.
\eee
Here, $E_{\nu_2}=E_i-E_f-E_{e_1}-E_{e_2}-E_{\nu_1}$ due to energy conservation. $E_i$, $E_f$, $E_{e_i}$
($E_{e_i}=\sqrt{p_{e_i}^2+m^2_e}$) and $E_{\nu_i}$ ($i=1,2$) are the energies of initial and final nuclei,
electrons and antineutrinos, respectively.
$F(Z_f, E_{e_i})$ denotes relativistic Fermi function and $Z_f=Z+2$.
${\cal A}^{2\nu}$ consists of products of nuclear matrix elements,
which depend on  lepton energies:
\bee \label{nucmat}
{\cal A}^{2\nu} &=&  g_V^4
\left[\frac{1}{4}|M^K_{F}+M^L_{F}|^2 + \frac{3}{4}|M^K_{F}-M^L_{F}|^2\right]\\
&& - g_V^2 g_A^2 \textrm{Re}\left\{ M^{K*}_{F} M^L_{GT} + M^{K*}_{GT} M^L_F \right\} \nonumber \\
&& + \frac{g_A^4}{3}
\left[\frac{3}{4}|M^K_{GT}+M^L_{GT}|^2+\frac{1}{4}|M^K_{GT}-M^L_{GT}|^2\right],
\nonumber
\eee
where
\bee \label{fagtmatt}
M^K_{F} &=& \sum_n \frac{K(0^+_n)}{2} F_n, ~~~
M^L_{F} = \sum_n \frac{L(0^+_n)}{2} F_n, \nonumber \\
M^K_{GT} &=& \sum_n \frac{K(1^+_n)}{2} G_n, ~~~
M^L_{GT} = \sum_n \frac{L(1^+_n)}{2} G_n, \nonumber \\
\eee
with
\bee
F_n &=&
\langle 0^+_f \parallel \sum_{m}\tau^-_m \parallel 0^+_{n}\rangle
\langle 0^+_n \parallel \sum_{m}\tau^-_m \parallel 0^+_{i} \rangle, \nonumber\\
G_n &=&
\langle 0^+_f \parallel \sum_{m}\tau^-_m \sigma_m \parallel 1^+_{n}\rangle
\langle 1^+_n \parallel \sum_{m}\tau^-_m \sigma_m \parallel 0^+_{i} \rangle, \nonumber\\
\eee
and energy denominators are
\bee
K_n(J^+)&=& ~~\frac{2}{(2E_n(J^+)-E_i-E_f)+\epsilon_{K}}\nonumber\\
&&+\frac{2}{(2E_n(J^+)-E_i-E_f)-\epsilon_{K}}\nonumber \\
L_n(J^+)&=& ~~\frac{2}{(2E_n(J^+)-E_i-E_f)+\epsilon_{L}}\nonumber\\
&&+\frac{2}{(2E_n(J^+)-E_i-E_f)-\epsilon_{L}}.\nonumber
\eee
Here, $|0^+_i\rangle$, $|0^+_f\rangle$ are the $0^+$ ground states of the initial and final even-even nuclei,
respectively, and $|0^+_n\rangle$ ($|1^+_n\rangle$) are all possible states of the intermediate nucleus
with angular momentum and parity $J^\pi = 0^+$ ($1^+$) and energies $E_n(0^+)$ ($E_n(1^+)$).
$\epsilon_{K}=E_{e_2}+E_{\nu_2}-E_{e_1}-E_{\nu_1}$ and $\epsilon_{L}=E_{e_1}+E_{\nu_2}-E_{e_2}-E_{\nu_1}$.
We note that formally in the limit $2E_n-E_i-E_f=0$  one ends up with ${\cal A}^{2\nu}=0$.
The maximal value of $|\epsilon_K|$ and $|\epsilon_L|$ is the $Q$ value of the process. For
$2\nu\beta\beta$ decay with energetically forbidden transition to intermediate nucleus
($E_n-E_i > - m_e$) the quantity $2E_n(J^+)-E_i-E_f = Q + 2 m_e + 2 (E_n-E_i)$ is always larger
than the $Q$ value. We clarify later that this quantity can be expressed as a combination of
residual interactions of nuclear Hamiltonian.

The calculation of the decay probability is usually simplified by an approximation
\begin{equation}
K_n(J^+) \sim L_n(J^+) \sim \frac{2}{E_n(J^+) - (E_i+E_f)/2}.
\end{equation}
Then we obtain
\bee \label{nucmatt}
{\cal A}^{2\nu} &=&  |g_V^2 M^{2\nu}_F - g_A^2 M^{2\nu}_{GT}|^2
\eee
with the Fermi and GT matrix elements given by
\bee \label{fagtmat}
&& M^{2\nu}_{F} =
\sum_n \frac{\langle 0^+_f \parallel T^-\parallel 0^+_{n}\rangle
\langle 0^+_n \parallel T^- \parallel 0^+_{i} \rangle}{E_n(0^+) - (E_i+E_f)/2},\nonumber\\
&& M^{2\nu}_{GT} = \nonumber\\
&& \sum_n \frac{\langle 0^+_f \parallel \sum_{m}\tau^-_m \sigma_m \parallel 1^+_{n}\rangle
\langle 1^+_n \parallel \sum_{m}\tau^-_m \sigma_m \parallel 0^+_{i} \rangle}
{E_n(1^+) - (E_i+E_f)/2}.\nonumber\\
\eee
Here, $T^- = \sum_{m} \tau^-_m$ is the total isospin-lowering operator.
As a result of the above approximation, the separation of phase space factor and
nuclear matrix elements is achieved.

The calculation of $M^{2\nu}_F$ and $M^{2\nu}_{GT}$ needs to evaluate explicitly the matrix
elements to and from the individual $|0^+_n\rangle$ and $|1^+_n\rangle$ states
in the intermediate odd-odd nucleus, respectively. In the shell model and
IBM calculation of these matrix elements the sum over virtual intermediate nuclear states
is completed by closure after replacing $E_n(J^+)$ by some average value
$\overline{E}_n(J^+)\rangle$:
\begin{eqnarray}
M^{2\nu}_{F} &=& \frac{M^{2\nu}_{F-cl}}{\overline{E}_n(0^+) - (E_i+E_f)/2},\nonumber\\
M^{2\nu}_{GT} &=& \frac{M^{2\nu}_{GT-cl}}{\overline{E}_n(1^+) - (E_i+E_f)/2}
\end{eqnarray}
with
\begin{eqnarray}
M^{2\nu}_{F-cl} &=& \langle 0^+_f| T^- T^-| 0^+_i\rangle, \nonumber\\
M^{2\nu}_{GT-cl}&=& \langle 0^+_f| \sum_{m,n} \tau^-_m \tau^-_n \vec{\sigma}_m\cdot\vec{\sigma}_n| 0^+_i\rangle.
\end{eqnarray}
The validity of the closure approximation is as good as the guess about the average energy
to be used. This approximation might be justified in the case where
there is a dominance of transition through a single state of the intermediate nucleus.

The $T^-$ operator connects states only  in the same isospin multiplet. $M^{2\nu}_{F,F-cl}$ is
non-zero only to that extent that Coulomb interaction mixes states of different multiplets.
As an example $2\nu\beta\beta$-decay transition ${^{48}{\rm Ca}} \rightarrow {^{48}{\rm Ti}}$
can be considered. The ground state of parent and daughter nuclei  can be identified
with $T=4$, $M_T=4$ and  $T=2$, $M_T=2$ states, respectively. A crude estimate of the mixing
of the $T=2$, $M_T=2$ state with $T = 4$ $M_T =2$ analog of the $^{48}$Ca ground state
due to the isotensor piece of Coulomb force implies a negligible small value $M^{2\nu}_{F-cl} < 0.02$
for this and some other $2\nu\beta\beta$-decay transitions \cite{hax84}.

The GT operator $\sum_n \tau^-_n \sigma_n$ connects states only within the same
spin-isospin multiplet of the SU(4) symmetry, which leads to new conserved
quantum numbers in addition to those of spin and isospin. The ground state of
the initial $(A,Z)$ even-even nucleus belongs to the multiplet [$n,n,0$] with
spin $S=0$ and isospin $T=n=(N-Z)/2$ and it is the only state of this nucleus
belonging to that multiplet. In the neighbor ($A,Z+1$) odd-odd nucleus there
are two states of the multiplet [$n,n,0$], namely the isobaric analog state
with $T=n$, $S=0$ and the GT state with $T=n-1$, $S=1$. In the final ($A,Z+2$) even-even
nucleus, the states belonging to the [$n,n,0$] multiplet  are the double isobaric
$T=n$, $S=0$ and two GT states with $T=n-2$, $S=0$ and $T=n-2$, $S=2$. The ground state
of the final ($A,Z+2$) even-even nucleus with $T=n-2$, $S=0$ belongs to the
multiplet [$n-2,n-2,0$]. The SU(4) limit results in vanishing
matrix elements $M^{2\nu}_{GT}$ and $M^{2\nu}_{GT-cl}$. The non zero double GT matrix element
requires a breaking of the SU(4) symmetry able to mix the ground and
excited states of the final nucleus. The dynamical origin of
breaking the SU(4) symmetry is associated with the spin-orbit and
the tensor potentials which affect mainly the mean field.
Another possibility is the difference between strength
triplet-singlet, triplet-triplet, and singlet-singlet channels
of the central potential. We show later that the
$2\nu\beta\beta$-decay NMEs does not depend explicitly on the
mean field part of the nuclear Hamiltonian. In contrast,
mainly the differences between triplet-singlet and singlet-triplet
(spin-isospin) interactions of nuclear Hamiltonian  contribute to
the $2\nu\beta\beta$ process. This small violation of the SU(4)
symmetry will be studied in an exactly solvable model within the
perturbation theory.

%
\section{Schematic Hamiltonian expressed with generators of SO(8) group}
%

We consider an exactly solvable model \cite{engel} with a set of degenerate single-particle orbitals,
characterized by $l$, $s=1/2$, and $t=1/2$. The total number of single-particle states is
$\Omega = \sum_l (2l+1)$. The model is made solvable by building a basis entirely
from $L=0$ operators, i.e., pairs of nucleons with spin $S=0$ and isospin $T=1$
and with $S=1$ and $T=0$ are allowed. The Hamiltonian of the model is an extension
of the Hamiltonian  introduced in Ref.\cite{engel} and possesses the main qualitative features
of a realistic Hamiltonian relevant to double-$\beta$ decay. It contains
proton and neutron single-particle terms and the two-body residual interaction, which components are isovector
spin-0, isoscalar spin-1 pairing and the particle-hole force in the $T=1$, $S=1$ channel. We have
\begin{widetext}
\bee \label{hamsopor}
H&=&\underbrace{e_n N_n+e_pN_p-g_{pair}\lf\sum_{M_T=-1,0,1}A_{0,1}\dg(0,M_T)A_{0,1}(0,M_T)
+\sum_{M_S=-1,0,1}A_{1,0}\dg(M_S,0)A_{1,0}(M_S,0) \rh+g_{ph}\sum_{a,b}E_{a,b}^{\dagger}E_{a,b}}_{H_0}\nonumber \\
&& +\underbrace{(g_{pair}-g_{pp}^{T=0})\sum_{M_S=-1,0,1}A_{1,0}\dg(M_S,0)A_{1,0}(M_S,0)
+(g_{pair}-g_{pp}^{T=1})A_{0,1}\dg(0,0)A_{0,1}(0,0)}_{H_I}.
\eee
\end{widetext}
Here, $g_{pair}$, $g_{pp}^{T=1}$,$g_{pp}^{T=0}$, and $g_{ph}$ denote the strengths of the isovector-like nucleon spin-0
pairing $(L=0,S=0,T=1,M_T\pm1)$, isovector proton-neutron spin-0 pairing $(L=0,S=0,T=1,M_T=0)$, isoscalar spin-1 pairing
$(L=0,S=1,T=0)$, and particle-hole force $(L=0,S=1,T=1)$, respectively. The proton number operator $N_p$,
neutron number operator $N_n$, particle-particle operators  $A_{S,T}\dg(M_S,M_T)$, and particle-hole GT
operators $E_{a,b}$ are defined in Appendix \ref{op}.

The six particle-particle operators $A_{S,T}\dg(M_S,M_T)$ and their Hermitian conjugates together with nine
particle-hole GT operators $E_{a,b}$,  total spin $\vec{S}$ and isospin $\vec{T}$ operators, and
total particle number operator (defined for convention as $Q_0=\Omega-\frac{1}{2}(N_p+N_n)$) represent
28 operators which generate the group SO(8) \cite{pang}. In case of seniority zero, which we will henceforth
assume, the SO(8) irreducible representation is specified by 7 numbers: (i) spatial degeneracy number of levels
$\Omega=\sum_l 2l+1$; (ii) eigenvalue of $Q_0$ operator $\lambda=\Omega-N/2$; (iii)
$n$ which corresponds to the irreducible SU(4) representation $[n,n,0]$;
(iv) total spin number $S$; (v) total spin projection $M_S$; (vi) total isospin number $T$; and  (vii)
total isospin projection $M_T$.  As we are constrained by the set of degenerate $l$ shells with
total degeneracy $\Omega$ and  given particle number $N$,
for basis state we introduce the abbreviation as follows:
\be \label{baza}
\ket{S,M_S,T,M_T,n},~\textrm{or}\quad \ket{S T n}.
\ee
We note that matrix elements of generators SO(8) group are known in this basis  \cite{pang,hechtpang},
which allows diagonalization of the Hamiltonian (\ref{hamsopor}).
Relevant expressions can be found in Apppendix \ref{ap}.

The physics associated with a simplified version of the Hamiltonian in Eq. (\ref{hamsopor})  was
studied previously with emphasis on energy levels \cite{pang,flow,evans}, the extreme sensitivity of $M^{2\nu}_{GT}$ to
the strength of proton-neutron particle-particle interaction \cite{vogel86}, and the interplay between
the isoscalar and isovector pairing models \cite{engel}. Here, we  discuss the role of the
violation of the isospin SU(2) and spin-isospin SU(4) symmetries and of different components
of Hamiltonian in the calculation of two-neutrino double-$\beta$ decay matrix elements by taking
the advantage of the perturbation theory.

The Hamiltonian in (\ref{hamsopor}) is decomposed in two parts: $H_0$, the unperturbed Hamiltonian,
and $H_I$, the perturbing one. The eigenstates of unpertubated Hamiltonian $H_0$
are characterized by the number of nucleon pairs (only systems with even
number of nucleons are considered), the isospin $T$, and a quantum number $n$
corresponding to the irreducible SU(4) representation [$n,n,0$]. The possible values of
quantum number $n$ for system with ${\cal N}$ particles in the set of
degenerate $l$ shells with degeneracy $\Omega$ and given $T$, $S$ are
$S+T,S+T+2, \cdots ,n_{max}$, where $n_{max}={\cal N}/2$ if ${\cal N}/2\leq\Omega$ and
$n_{max}=2\Omega-{\cal N}/2$ otherwise \cite{engel}.
The single-particle and particle-hole interaction components of $H_0$ violate both isospin
and spin-isospin symmetries and as a consequence energies of states
with the same quantum numbers $T$ and $n$ are different for a given $T_z=(N-Z)/2$ ($T_z\equiv M_T$).
$N$ and $Z$ are numbers of neutrons and protons, respectively.
If $g_{pair}=g_{pp}^{T=0}$ and $g_{pair}=g_{pp}^{T=1}$ the isospin and spin-isospin symmetries
of particle-particle interaction are restored we get $H=H_0$ and $H_I=0$.
If $g_{pair} \neq g_{pp}^{T=0}$ and/or $g_{pair} \neq g_{pp}^{T=1}$, the Hamiltonian (\ref{hamsopor})
is not more diagonal in basis (\ref{baza}) and states  with different quantum numbers  $T$ and $n$
are mixed. The eigenstate of the Hamiltonian $\ket{S,M_S,T',M_T,n'}$
can be expressed with eigenstates of unperturbated Hamiltonian $H_0$ as follows:
\bee
&\ket{S,M_S,T',M_T,n'}= \nonumber \\
&\sum_{n,T} c^{T'n'}_{S,M_S,T,M_T,n}\ket{S,M_S,T,M_T,n}.
\eee
Here, we assume a small violation of the SU(4) symmetry, which can be treated by a
perturbation theory. The prime symbol by quantum numbers $T$ and $n$  ($T'$ and $n'$) indicates that  these quantum numbers are not more 
   good quantum numbers due to the violation of SU(4) symmetry
   and that the dominant component in the expansion over states with a good isospin and
the SU(4) quantum number is that with $T'=T$ and $n'=n$.
A diagonalization of Hamiltonian requires calculation of matrix elements
\bee
&&\bra{S,M_S,T,M_T,n\pm 2} H \ket{S,M_S,T,M_T,n},\nonumber\\
&& \bra{S,M_S,T\pm 2,M_T,n} H \ket{S,M_S,T,M_T,n},\nonumber\\
&& \bra{S,M_S,T\pm 2,M_T,n\pm 2} H \ket{S,M_S,T,M_T,n}.\nonumber
\eee
The corresponding expressions are given explicitly in Apppendix \ref{ap}.

We shall assume a small violation of the SU(4) symmetry due to $H_I\neq 0$,
namely $g_{pair} \simeq g_{pp}^{T=0}$ and/or $g_{pair} \simeq g_{pp}^{T=1}$.
For the numerical application we consider a set of parameters as follows \cite{engel}:
\bee\label{parameters}
&& e_p = 1.2 \textrm{MeV}\qquad e_n = 1.1\textrm{MeV}\qquad\Omega=12, \nonumber \\
&& {\cal N} = 20,\qquad  g_{pair} = 0.5 \textrm{MeV},\qquad    g_{ph} = 1.5~ g_{pair}.\qquad        \nonumber \\
\eee
The initial, intermediate and final states of the double-beta decay transition
will be identified with the isospin projection $T_z =4$, 3, and 2. For these
three values of $T_z$ the corresponding numbers of neutrons
and protons ($N,Z$) are  (14,6), (15,7), and (12,8), respectively.


\begin{figure}[!t]
\centering
\includegraphics[width=1.1\columnwidth]{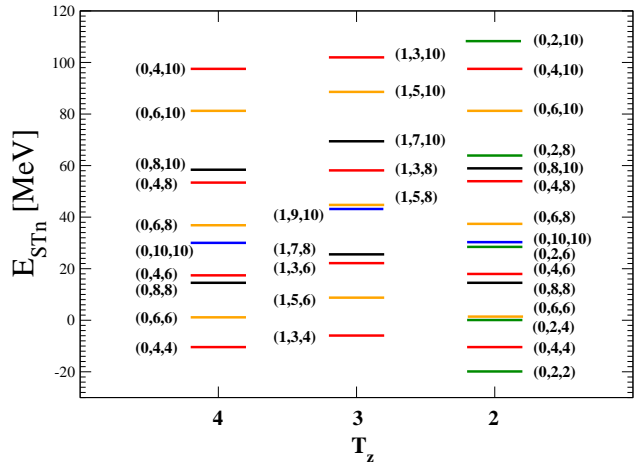}
\caption{(Color online) Eigenenergies $E_{STn}$ of the Hamiltonian (\ref{hamsopor})
for set of parameters (\ref{parameters}), $T_z$=4, 3, and 2 and by assuming $H_I=0$.
Energy states are labeled by spin, isospin, and the SU(4)
quantum number $n$: ($S,T,n$). The levels with different  value of  $S+T$ are displayed
in different colors online: $S+T= 2$ (green), 4 (red), 6 (orange), 8 (black), and 10 (blue).
\label{elevs}}
\end{figure}


In Fig. \ref{elevs} we present states with energy $E_{S T n}$ of different isotopes.
The considered level scheme illustrates the situation with double GT transition for
${^{48}}$Ca  as the isospin and its projection of the initial and final ground states
correspond to those of $^{48}$Ca and ${^{48}}$Ti.
We note that in nuclear physics the isospin symmetry is conserved to a great extent.
Within the studied model in the SU(4) symmetry limit the ground states of ${^{48}}$Ca
and ${^{48}}$Ti can be identified  with $S=0$, $T=4$, $T_z=4$, $n=4$, and $S=0$, $T=2$, $T_z=2$, $n=2$,
respectively, and the intermediate states in ${^{48}}$Sc with S=1 (T=3,5,7, and 9)
T$_z$=3 (n=4, 6, 8 and 10). As the GT operator is not changing quantum number
$n$, the double GT matrix elements  connecting initial and final ground states
is nonzero only to the extent the breaking of SU(4) symmetry mixes the high-lying
(0,4,4) analog of the ${^{48}}$Ca ground state into (0,2,2) analog of the ${^{48}}$Ti.

%
\section{Double Fermi and GT matrix elements within the perturbation theory}
%

We study the double GT and Fermi matrix elements using perturbation theory within the discussed model
close to a point of restoration of the SU(4) symmetry of particle-particle interaction of $H$.
First, we assume a conservation of the isospin symmetry by the particle-particle interaction
and a subject of interest will be $M^{2\nu}_{GT}$ as function of the isospin of the initial state.
Then a weak violation of the isospin symmetry is allowed and the dependence of  $M^{2\nu}_F$ and
$M^{2\nu}_{GT}$ on both quantities $g_{pair} - g_{pp}^{T=0}$ and $g_{pair} - g_{pp}^{T=1}$,
which violates the SU(4) symmetry, is analyzed.

%
\subsection{The GT matrix element in the case of isospin symmetry}
%

We consider a small violation of the SU(4) spin-isospin symmetry in
nuclear Hamiltonian (\ref{hamsopor}) due to $g_{pair}\neq g_{pp}^{T=0}$
and that isospin is a good quantum number, i.e., $g_{pair} = g_{pp}^{T=1}$,
which implies $M^{2\nu}_F=0.$

As an example we discuss in details the GT matrix element for $2\nu\beta\beta$ decay from the state
with $S=0$, $T=M_T=4$ to the state with $S=0$, $T=M_T=2$. The corresponding transition is
\be \label{proc}
\ket{0,0,4,4,4}\rightarrow \ket{1,M_S,3,3,4} \rightarrow \ket{0,0,2,2,2}.
\ee
In the case $g_{pair} = g_{pp}^{T=0}$ one finds that $M^{2\nu}_{GT}=0$ as eigenstates
of the GT operators are diagonal in SU(4) quantum number  $n$ and the initial and final
states are assigned into different SU(4) multiplets. By breaking the SU(4) symmetry
of particle-particle interaction the quantum number $n$ is not more a good quantum number
and states with different $n$
are mixed. By keeping in mind a small violation of this symmetry we denote
perturbated states and their energies with a superscript prime symbol ($|S,M_S,T,M_T,n' \rangle$, $E'_{S,M_S,T,M_T,n}$),
unlike the states with a definite quantum number $n$ ($|S,M_S,T, M_T,n\rangle$, $E_{S,M_S,T, M_T, n}$).

Up to the second order of parameter $(g_{pair}-g_{pp}^{T=0})$ we get (for sake of simplicity a shorter
notation of states and energies (\ref{baza}) is used)
\bee \label{efi}
E'_{022}&=&E_{022}+\bra{022}H_I\ket{022}+\frac{|\bra{024}H_I\ket{022}|^2}{E_{022}-E_{024}}\nonumber \\
&=&12e_n+8e_p-94g_{pair}+6g_{ph}\nonumber \\
&+&(g_{pair}-g_{pp}^{T=0})\frac{132}{5}- \frac{(g_{pair}-g_{pp}^{T=0})^2}{10g_{pair}+20g_{ph}}\frac{8316}{25}\nonumber \\
\eee
\bee  \label{eme}
E'_{134}&=&E_{134}+\bra{134}H_I\ket{134}+\frac{|\bra{136}H_I\ket{134}|^2}{E_{134}-E_{136}}\nonumber \\
&=&13e_n+7e_p-84g_{pair}+18g_{ph}\nonumber \\
&+&(g_{pair}-g_{pp}^{T=0})\frac{201}{7} -\frac{(g_{pair}-g_{pp}^{T=0})^2}{14g_{pair}+28g_{ph}}\frac{10125}{49}\nonumber \\
\eee
\bee \label{ein}
E'_{044}&=&E_{044}+\bra{044}H_I\ket{044}+\frac{|\bra{046}H_I\ket{044}|^2}{E_{044}-E_{046}}\nonumber \\
&=&14e_n+6e_p-84g_{pair}+12g_{ph} \nonumber \\
&+&(g_{pair}-g_{pp}^{T=0})\frac{108}{7}-\frac{(g_{pair}-g_{pp}^{T=0})^2}{14g_{pair}+28g_{ph}}\frac{7425}{49}.\nonumber \\
\eee


\begin{figure}[!t]
\centering
\includegraphics[width=1.1\columnwidth]{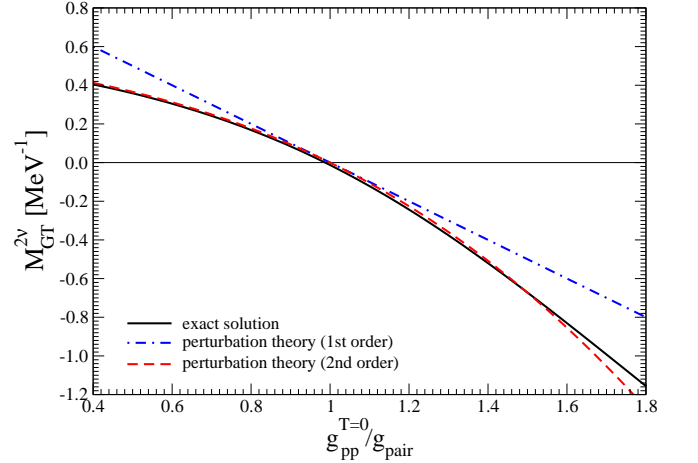}
\caption{ (Color online) Matrix element $M^{2\nu}_{GT}$ for the double-GT two-neutrino double-$\beta$ decay mode
as function of ratio  $g_{pp}^{T=0}/g_{pair}$  for a set of parameters (\ref{parameters}). Exact results are indicated
with a solid line. The results obtained within the perturbation theory up to the first and second
order in $H_I$ contribution to Hamiltonian (\ref{hamsopor}) are shown with dashed-dotted and dashed lines, respectively.
The restoration of spin-isospin symmetry of particle-particle interaction is achieved for $g_{pp}^{T=0}/g_{pair}=1$.
\label{gtplot}}
\end{figure}


For the double GT matrix element we have
\begin{eqnarray}\label{fermione}
M^{2\nu}_{GT} &=& \sum_{n'}^{}\frac{ \bra{0 2 2'}\vec{\sigma}\tau^-\ket{1 3 n'}\cdot \bra{1 3 n'}\vec{\sigma} \tau^-\ket{0 4 4'}}
{E'_{1 3 n}-\lf E'_{0 4 4}-E'_{0 2 2}\rh/2}, \nonumber\\
&\simeq& \frac{ \bra{0 2 2'}\vec{\sigma}\tau^-\ket{1 3 4'}\cdot \bra{1 3 4'}\vec{\sigma} \tau^-\ket{0 4 4'}}
{E'_{1 3 4}-\lf E'_{0 4 4}-E'_{0 2 2}\rh/2}.
\label{medfermi}
\end{eqnarray}
The allowed intermediate states $\ket{1 3 n'}$ are those with $S=1$, $T=3$ and $n'=4,6,8$ and $10$. We note that
up to second order of perturbation theory there is only a single contribution through the intermediate state
$\ket{1 3 4'}$ and the product of two corresponding $\beta$-amplitudes
(numerator of (\ref{medfermi}) takes the form
\begin{eqnarray}
&&\left\langle 0 2 2'\left|\vec{\sigma}\tau^-\right|1 3 4'\right\rangle\cdot \left\langle 1 3 4'\left| \vec{\sigma}\tau^-\right|0 4 4'\right\rangle
=144\sqrt{\frac{231}{35}}\nonumber\\
&&\times\left(\frac{(g_{pair}-g_{pp}^{T=0})}{10g_{pair}+20g_{ph}}-\frac{267(g_{pair}-g_{pp}^{T=0})^2}{35(10g_{pair}+20g_{ph})^2}\right).
\nonumber\\
\end{eqnarray}
We see that if $g_{pair}=g_{pp}^{T=0}$ GT matrix element vanishes.
With help of Eqs. (\ref{efi}), (\ref{eme}) and (\ref{ein}) for the energy denominator in (\ref{medfermi})
we obtain
\begin{eqnarray} \label{denominator}
&&(2E'_{1 3 4}-E'_{0 2 2}-E'_{0 4 4})/2=5g_{pair}+9g_{ph}\nonumber\\
&&+(g_{pair}-g_{pp}^{T=0})\frac{39}{5}+\frac{(g_{pair}-g_{pp}^{T=0})^2}{g_{pair}+2g_{ph}}\left(\frac{1249263}{171500} \right).
\nonumber\\
\end{eqnarray}
It is worth noting that neither the numerator nor denominator of $M^{2\nu}_{GT}$ depend explicitly on the
single-particle energies $e_n$ and $e_p$. If we restrict our consideration
to the first-order perturbation theory we find
\begin{eqnarray}
M^{2\nu}_{GT} &=& \frac{144\sqrt{\frac{231}{35}}(g_{pair}-g_{pp}^{T=0})}
{(5g_{pair}+9g_{ph})(10g_{pair}+20g_{ph})}.
\end{eqnarray}

In Fig. \ref{gtplot} $M^{2\nu}_{GT}$ is plotted as function of ratio $g_{pp}^{T=0}/g_{pair}$.
We see that results obtained within the second-order perturbation theory agree well with exact results within
a large range of this parameter. For $g_{pp}^{T=0}/g_{pair}=1$ the restoration of the SU(4) symmetry
of particle-particle interaction is achieved,
i.e.,  $M^{2\nu}_{GT}$ is equal to zero.
We notice that if the quantity  $g_{pp}^{T=0}/g_{pair}$ is within the range (0.8,1.2) the first-order perturbation
theory seems to be sufficient.

\begin{figure}[!t]
\centering
\includegraphics[width=1.1\columnwidth]{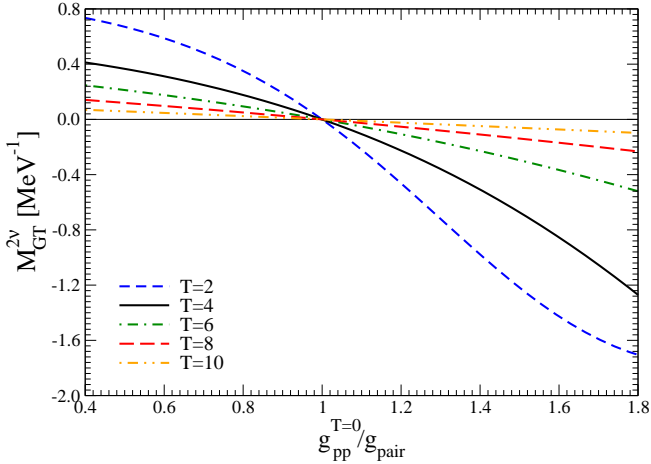}
\caption{ (Color online) Matrix element $M^{2\nu}_{GT}$
for the double-GT two-neutrino double-$\beta$ decay mode
as function of ratio  $g_{pp}^{T=0}/g_{pair}$  for different initial state with $T=M_T$ ($M_T=2$, 4, 6, 8,
and 10).
\label{gtplotrozne}}
\end{figure}

Usually, ground states of stable even-even nuclei are identified with isospin $T=T_z$.  The dependence of
$M^{2\nu}_{GT}$ on the isospin of the initial nucleus is presented in Fig. \ref{gtplotrozne}. We see that
for a fixed value of $g_{pp}^{T=0}/g_{pair}$ (i.e., breaking of the SU(4) symmetry) the absolute value
of $M^{2\nu}_{GT}$ decreases with increasing isospin $T$. We note that apart from the shell effects of magic nuclei
this tendency is observed also for measured $2\nu\beta\beta$-decay matrix elements \cite{barabash}.


%
\subsection{The Fermi and GT matrix elements in the case of broken SU(2) and SU(4) symmetries}
%

The main task to be addressed in this subsection is determining the dependence of $M^{2\nu}_{F}$
and $M^{2\nu}_{GT}$ on both quantities $g_{pp}^{T=1}/g_{pair}$ and $g_{pp}^{T=0}/g_{pair}$.
Recall that $g_{pp}^{T=1} \neq g_{pair}$ breaks both the SU(2) isospin and the SU(4) spin-isospin symmetries
of particle-particle interaction
unlike $g_{pp}^{T=0} \neq g_{pair}$, which is associated only with the violation of the SU(4) symmetry.


\begin{figure}[!t]
\centering
\includegraphics[width=1.1\columnwidth]{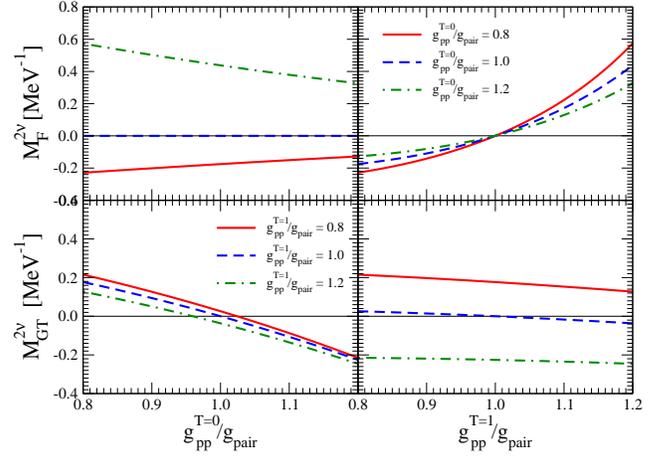}
\caption{ (Color online) Matrix elements $M^{2\nu}_{F}$ and $M^{2\nu}_{GT}$ as function 
of ratios  $g_{pp}^{T=0}/g_{pair}$  and $g_{pp}^{T=1}/g_{pair}$
for transition from the initial $|0 4' 4'\rangle$ to final $|0 2' 2'\rangle$
ground state and a set of parameters (\ref{parameters}).
The results are obtained within the perturbation theory up to the second
order.
\label{Fig.4}}
\end{figure}

We consider the $2\nu\beta\beta$-decay transition from the initial
$|0 4' 4'\rangle$ to final $|0 2' 2'\rangle$  ground state. Up to the first
order in the perturbation theory for double Fermi and GT matrix
elements we find
\begin{eqnarray}
M^{2\nu}_{F}
&=&- \frac{48\sqrt{\frac{33}{5}}\left(g_{pair}-g_{pp}^{T=1}\right)}
{(5g_{pair}+3g_{ph})(10g_{pair}+6g_{ph})},\\
M^{2\nu}_{GT}&=& \frac{144\sqrt{\frac{33}{5}}}
{5g_{pair}+9g_{ph}}\left\{\frac{(g_{pair}-g_{pp}^{T=0})}{(10g_{pair}+20g_{ph})} \right.\nonumber\\
&&+  \left.\frac{2g_{ph}(g_{pair}-g_{pp}^{T=1})}
{(10g_{pair}+20g_{ph})(10g_{pair}+6g_{ph})}\right\}.\nonumber\\
\end{eqnarray}
We see that $M^{2\nu}_{F}$ depends only  on strength of the isovector interaction
$g_{pp}^{T=1}$ unlike $M^{2\nu}_{GT}$, which depends also on the strength of the
isoscalar interaction   $g_{pp}^{T=0}$. Due to the  violation of the isospin
symmetry the final ground state $|0 2' 2'\rangle$ is mixed with both
first $|0 4' 4'\rangle$  and second $|0 2' 4'\rangle$ excited states (see Fig. \ref{elevs}),
resulting in $g_{pp}^{T=1}$ contribution to $M^{2\nu}_{GT}$.

In Fig. \ref{Fig.4} we present behavior of $M^{2\nu}_{F}$ and $M^{2\nu}_{GT}$
as function of $g_{pp}^{T=1}/g_{pair}$ ($g_{pp}^{T=0}/g_{pair}$) for a particular
values of $g_{pp}^{T=0}/g_{pair}$ ($g_{pp}^{T=1}/g_{pair}$). Results were obtained
within the perturbation theory up to the second order. We see clearly that
for $g_{pp}^{T=1}/g_{pair}=0$ matrix element $M^{2\nu}_{F}$ does not depend on $g_{pp}^{T=0}$
and  varies strongly with change of $g_{pp}^{T=1}$.  A different
behavior offers $M^{2\nu}_{GT}$, which weakly depends on the $g_{pp}^{T=1}$ and
significantly on the $g_{pp}^{T=0}$. These conclusions agree qualitatively well
with results obtained for two-neutrino double-$\beta$ decay transitions
within the proton-neutron QRPA with restoration of the isospin
symmetry \cite{newpar13}. The advantage of the study which considered the
schematic model and in perturbation theory is that explicit dependence
of $M^{2\nu}_{F}$ and $M^{2\nu}_{GT}$ on isoscalar and isovector strength
of particle-particle interactions can be determined.

%
\section{Energy-weighted sum rule of $\Delta Z=2$ nuclei}
%

In Ref. \cite{dusan13} the double Fermi and GT sum rules associated with
$\Delta Z=2$ nuclei were introduced. We have
\begin{eqnarray} \label{energysum}
S^{ew}_{F}(i,f) &\equiv& \sum_{n}(E_n - \frac{E_i + E_f}{2})\bra{f} T^- \ket{n}\bra{n} T^- \ket{i}
\nonumber\\
&=& \frac{1}{2} \bra{f}\left[ T^-,\left[H, T^- \right]\right]\ket{i}\\
S^{ew}_{GT}(i,f)
&\equiv& \sum_{n}(E_n - \frac{E_i + E_f}{2})\bra{f} \vec{\cal O}_{GT} \ket{n}\cdot
\bra{n} \vec{\cal O}_{GT}\ket{i}
\nonumber\\
&=& \sum_M (-1)^M \frac{1}{2} \bra{f}\left[ ({\cal O}_{GT})_{-M},\left[H, ({\cal O}_{GT})_M\right]\right]\ket{i}
\nonumber\\
\end{eqnarray}
with
\begin{eqnarray}
\vec{\cal O}_{GT} = \sum_{k=1}^A \tau^-_k \mathbf{\vec{\sigma}_k},
\end{eqnarray}
where $|i>$ ($|f>$) are $0^+$ ground states of the initial (final)
even-even nuclei with energy $E_i$ ($E_f$),
and $|1^+_n>$ ($|0^+_n>$) are the $1^+$ ($0^+$) states in the
intermediate odd-odd nucleus  with energies $E_n$.

If there is a dominance of contribution through a single or few states
of the intermediate nucleus, energy-weighted sum rules (\ref{energysum})
might be exploited to adjust the strengths of the residual interaction
of Hamiltonian for nuclear structure calculations. The  left-hand side
of Eq. (\ref{energysum}) might be determined phenomenologically, unlike
the right-hand side of Eq. (\ref{energysum}), which requires evaluation
of the double commutator within a nuclear model
and can be expressed in terms of the strengths of residual interaction.
Due to a double commutator of nuclear Hamiltonian with charge-changing
Fermi and GT operators connecting states with $\Delta Z =2$
the energy-weighted sum rule $S^{ew}_{F,GT}(i,f)$
does not depend explicitly on the mean field part of the nuclear Hamiltonian.

\begin{table}[h]
   \caption{The coefficients $a$, $b$, $c$ and $d$ of the expansion of the energy denominator
$E'_{n} - \frac{E'_{i} + E'_{f}}{2}$ [see Eq. (\ref{denparam})] associated with the dominant
double GT (double Fermi)  transition from the initial ground state $|0 T=M_T n\rangle$
to the final ground state $|0~T=M_T-2~n-2\rangle$ through a single state of the intermediate nucleus
$|1 M_T~n\rangle$ ($|0~M_T~n\rangle$). Coefficients are presented for different isospin $T=M_T$ of the initial
state. \label{tch}}
   \begin{tabular}{lccccccc}
    \hline
$T=M_T$  & & Transition  & \multicolumn{4}{c}{Coefficients}\\\cline{4-7}
  & & & $a$& $b$ & $c$ & $d$  \\
\hline
\hline
2 &  & GT & 3 & 5 & $-59/15$ & $44/5$ \\
  &   & Fermi & 3 & 3 & $50/3$ & $-59/5$ \\
\hline
4 & & GT & 5 & 9 & $-64/35$ & $39/5$ \\
4 & & Fermi & 5 & 3 & $401/35$ & -192/35 \\
\hline
6 & & GT & 7 & 13 & $-71/63$ & $340/63$ \\
 &  & Fermi & 7 & 3 & $482/63$ & $-71/21$ \\
\hline
8 & & GT & 9 & 17 & $-80/99$ & $103/33$ \\
 & & Fermi & 9 & 3 & $469/99$ & $-80/33$ \\
\hline
10 & & GT & 11 & 21 & $-7/11$ & $12/11$ \\
   & & Fermi & 11 & 3 & $26/11$ & $-21/11$ \\
\hline
\end{tabular}
      \end{table}

We analyze $S^{ew}_{F,GT}$ for the Hamiltonian (\ref{hamsopor})
and by exploiting the commutation relations of the SO(8) group.
For the case $|i\rangle = |0 4 4'\rangle$,  $|f\rangle = |0 2 2'\rangle$,
we get
\begin{eqnarray} \label{hoo_F}
&& S^{ew}_{GT}(04'4',02'2')
\equiv\sum_{n}( E'_{13n} - \frac{E'_{044} + E'_{022}}{2})
\nonumber\\
&&~~~~~~~~~~~~~~~~~~~\times \bra{02'2'} \vec{\sigma}\tau^- \ket{13'n'}\cdot\bra{13'n'} \vec{\sigma}\tau^- \ket{04'4'}\nonumber\\
&&=6(g_{pp}^{T=0}-g_{pair})\bra{02'2'}A_{0,1}^{\dg}(0,-1)A_{0,1}(0,1)\ket{04'4'} \nonumber\\
&&-g_{ph}\bra{02'2'}\vec{\sigma}\tau^- \cdot \vec{\sigma}\tau^- \ket{04'4'}
-3 g_{ph} \bra{02'2'}T^-T^-\ket{04'4'}
\nonumber\\
\end{eqnarray}
and
\begin{eqnarray} \label{hoo_GT}
&& S^{ew}_{F}(04'4',02'2')
\equiv \sum_{n}( E'_{0,0,4,3,n} - \frac{E'_{044} + E'_{022}}{2})
\nonumber\\
&&~~~~~~~~~~~~~~~~~~~\times \bra{02'2'} T^-  \ket{04'n'}\cdot\bra{04'n'} T^- \ket{04'4'}\nonumber\\
&&=2(g_{pair}-g^{T=1}_{pp})\bra{02'2'}A_{0,1}^{\dg}(0,-1)A_{0,1}(0,1)\ket{04'4'}.\nonumber\\
\end{eqnarray}
We note that the dominant contribution to $S^{ew}_{GT}(044',022')$ and  $S^{ew}_{F}(044',022')$
comes from the transition through the single intermediate states $|134'\rangle$ and $|044'\rangle$,
respectively. By exploiting the first-order perturbation theory to evaluation of matrix elements
in Eqs. (\ref{hoo_GT}) and  (\ref{hoo_F}) for a combination of energies of involved states
we find
\bee
&& E'_{134} - \frac{E'_{044} + E'_{022}}{2} = 5g_{pair}+9g_{ph} \\ \nonumber
&&~~~~~~~~ - \frac{64}{35}(g_{pair}-g^{T=1}_{pp})+\frac{39}{5}(g_{pair}-g^{T=0}_{pp})
\label{den_GT}\\
&& E'_{0,0,4,3,4} -\frac{ E'_{044} + E'_{022}}{2} = 5g_{pair}+3g_{ph} \\ \nonumber
&&~~~~~+\frac{401}{35}(g_{pair}-g^{T=1}_{pp})-\frac{192}{35}(g_{pair}-g^{T=0}_{pp}).
\label{den_F}
\eee
The result in Eq. (\ref{den_GT}) is in agreement with above calculated expression for energy denominator
in Eq. (\ref{denominator}), which was derived by assumption of the isospin conservation.

We see that considered energy-weighted sum rules imply useful relations between three energies of states
appearing in the denominator of the double GT or Fermi matrix elements and nucleon-nucleon interactions.
The energy denominator associated with the dominant double-GT (double-Fermi) transition from the initial
ground state $|0 T=M_T n\rangle$ to the final ground state $|0~T=M_T-2~n-2\rangle$ through a single
state of the intermediate nucleus $|1 M_T~n\rangle$ ($|0~M_T~n\rangle$) can be written as
\bee\label{denparam}
&&E'_{n} - \frac{E'_{i} + E'_{f}}{2} =  \\ \nonumber
&& = a g_{pair}+ b g_{ph}+c\lf g_{pair} - g_{pp}^{T=1}\rh +d \lf g_{pair}-g_{pp}^{T=0} \rh.
\eee
Here, $a$, $b$, $c$ and $d$ are coefficients. The perturbation theory up to the first order
is considered. For different isospin $T=M_T$ of the initial ground-state
coefficients $a$, $b$, $c$ and $d$ are presented in Table \ref{tch}.  We see that for larger value
of $T$ the value of the energy denominator is affected less by the violation of both
the isospin and spin-isospin symmetries as it is for smaller value of $T$.

%
\section{Conclusions}
%

The anatomy of the two-neutrino double-$\beta$ decay matrix elements was studied within
a schematic model, which can be solved exactly and yet contains most of the qualitative
features of a more realistic description, and by taking  advantage of the perturbation
theory. We paid attention to violation  of both spin-isospin SU(4) and
isospin SU(2) symmetries of particle-particle interaction of Hamiltonian.
The isospin violation originates from the difference of
the proton-proton and the neutron-neutron pairing force compared to
the proton-neutron isovector pairing force. The break down of the SU(4) symmetry
is a consequence  of the difference of the like-nucleon pairing interaction compared to
the proton-neutron isoscalar interaction and/or to the proton-neutron
isovector interaction, which violates also the isospin symmetry.

By using perturbation theory, an explicit dependence  of the two-neutrino double-$\beta$ decay matrix elements
on different constituents of the Hamiltonian was established. It was found that the mean-field part of
the Hamiltonian does not enter explicitly in the decomposition of $M^{2\nu}_F$ and $M^{2\nu}_{GT}$ and is related only
to the calculation of unperturbated states of the Hamiltonian. This general conclusion is valid for any
mean field approximation. In the case of medium and heavy heavy nuclei the SU(4) symmetry
is strongly broken by the spin-orbit splitting, affecting strongly the mean field part, unlike the interaction
part of nuclear Hamiltonian. This fact might be an explanation for a smallness of $M^{2\nu}_{GT}$ being
governed by a small violation of the SU(4) symmetry by the particle-particle  interaction of the Hamiltonian.

The obtained expressions for $M^{2\nu}_{F}$ and $M^{2\nu}_{GT}$ supported by numerical
calculation up to the second order in perturbation theory  confirm the finding achieved within the
proton-neutron QRPA approach that $M^{2\nu}_F$ depends strongly on the isovector part of the
particle-particle neutron-proton interaction, unlike $M^{2\nu}_{GT}$, which depends strongly on
its isoscalar part. By assuming a fixed violation of the SU(4) symmetry by the particle-particle
interaction it is shown that value  of $M^{2\nu}_{GT}$
decreases by an increase of the isospin of the initial ground state. This tendency is
found also  in the case of measured double-GT matrix elements being partially spoiled by
different pairing properties. We also showed that $M^{2\nu}_{GT}$ contains a small
component due to violation of the isospin symmetry. By keeping in mind that
in nuclear physics the isospin symmetry is conserved to a great extent it is recommended
for evaluation of double-$\beta$ decay matrix elements to use many-body approaches with a
conservation or restoration of the isospin symmetry \cite{newpar13,iachello15}.

An important result coming from the analysis within
perturbation theory is that  there is a dominance of double-$\beta$ decay transition
through a single intermediate state. Further, the importance of the energy-weighted sum rule
associated with $\Delta Z = 2$ nuclei for fitting different components of the residual interaction of
the Hamiltonian was pointed out. It goes without saying that further studies, in particular those
in which a realistic nuclear Hamiltonian is used, are of great interest.

\acknowledgments
The authors thank  K. Muto for useful discussions.
This work is supported in part by the VEGA Grant Agency
of the Slovak Republic under Contract No. 1/0876/12 and by the Ministry
of Education, Youth and Sports of the Czech Republic under Contract No. LM2011027.

\begin{widetext}
%
\appendix
\section{Operators in SO(8) schematic model \label{op}}
%

We consider a set of single-particle states with the associated creation and annihilation
operators, $a\dg_{l m m_s m_t}$ and $a_{l m m_s m_t}$, which are  labeled by orbital angular
momentum $l$, its projection on $z$ axis $m$, and $z$ components of spin ($m_s=1/2$) and  isospin ($m_t=1/2$).

The particle-particle operators entering the  Hamiltonian (\ref{hamsopor}) are given by \cite{pang}
\bee \label{coperst}
A_{S,T}\dg(M_S,M_T)=\sum_{l,m,m_s,m_t}\sqrt{l+\frac{1}{2}}
C_{lmlm^{\prime}}^{00}
C_{\frac{1}{2}m_t\frac{1}{2}m_t^{\prime}}^{TM_T}C_{\frac{1}{2}m_s\frac{1}{2}m_s^{\prime}}^{SM_S}a\dg_{lmm_sm_t}a\dg_{lm^{\prime}m^{\prime}_sm^{\prime}_t},\quad (S,T)=(0,1)~\textrm{or}~(1,0),
\eee
the particle-hole GT operators take the form
\bee \label{gtoperst}
E_{a,b}=\sum_{l,m,m_s,m_t}\bra{(m_s+a)(m_t+b)}\sigma_a \tau_b \ket{m_s m_t}a\dg_{lm(m_s+a)(m_t+b)}a^{\phantom{\dagger}}_{lmm_sm_t},
\eee
and particle number operators are written as
\bee \label{numberoperators}
N_{i}=\sum_{l,m,m_s,m,l} a\dg_{lm m_s {m_t}_i}a^{\phantom{\dagger}}_{lmm_s{{m_t}_i}}, \quad i=p,n,  \quad m_{t_n,t_p}=\pm 1/2.
\eee
Here, $\sigma_a$ and $\tau_b$ represent spherical components of the single-particle Pauli spin and isospin operators
with convention used in Ref.\cite{pang}.

%

\section{Relevant SO(8) matrix elements \label{ap}}
%

The matrix elements of SO(8) operators in the basis of zero-seniority states are given in Refs. \cite{pang}
and \cite{hechtpang}. Here, we present the SO(8) matrix elements relevant for calculation
of the double GT and Fermi matrix elements. We have
\bee
&\bra{S,M_S,T ,M_T, n}\sum_{M_S}A\dg_{1,0}(M_S,0)A_{1,0}(M_S,0)\ket{S,M_S,T, M_T, n} \nonumber \\
&=\frac{(\Omega+n+\lambda+6)(\Omega-n-\lambda)}{8(n+2)(n+3)}\left[\frac{(S+1)(n+S+T+4)(n+S-T+3)+S(n-S+T+3)(n-S-T+2)}{(2S+1)}\right] \nonumber \\
&+\frac{(\Omega-n+\lambda+2)(\Omega+n-\lambda+4)}{8(n+1)(n+2)}\left[\frac{(S+1)(n-S+T+1)(n-S-T)+S(n+S-T+1)(n+S+T+2)}{(2S+1)}\right]   \nonumber
\eee
\bee
&\bra{S,M_S,T ,M_T, n}\sum_{M_T}A\dg_{0,1}(0,M_T)A_{0,1}(0,M_T)\ket{S,M_S,T, M_T, n} \nonumber \\
&=\frac{(\Omega+n+\lambda+6)(\Omega-n-\lambda)}{8(n+2)(n+3)}\left[\frac{(T+1)(n+T+S+4)(n+T-S+3)+T(n-T+S+3)(n-T-S+2)}{(2T+1)}\right] \nonumber \\
&+\frac{(\Omega-n+\lambda+2)(\Omega+n-\lambda+4)}{8(n+1)(n+2)}\left[\frac{(T+1)(n-T+S+1)(n-T-S)+T(n+T-S+1)(n+T+S+2)}{(2T+1)}\right]   \nonumber \nonumber
\eee
\bee
&\bra{S,M_S,T ,M_T, n+2}\sum_{M_T}A\dg_{0,1}(0,M_T)A_{0,1}(0,M_T)\ket{S,M_S,T, M_T, n} \nonumber \\
&=-\bra{S,M_S,T ,M_T, n+2}\sum_{M_S}A\dg_{1,0}(M_S,0)A_{1,0}(M_S,0)\ket{S,M_S,T, M_T, n} \nonumber \\
&=\frac{1}{8(n+3)}\sqrt{\frac{(\Omega+n+\lambda+6)(\Omega+n-\lambda+6)(\Omega-n+\lambda)(\Omega-n-\lambda)}{(n+2)(n+4)}}\nonumber \\
&\times \sqrt{(n+S+T+4)(n+S-T+3)(n-S+T+3)(n-S-T+2)}
\nonumber \eee
\bee
&\bra{S,M_S,T ,M_T, n}\sum_{a,b}E_{ab}^{\dagger}E_{ab}\ket{S,M_S,T, M_T, n} \nonumber \\
&=n(n+4)-S(S+1)-T(T+1)
\nonumber
\eee
\bee
&\bra{S,M_S,T ,M_T, n} A\dg_{0,1}(0,0)A_{0,1}(0,0)\ket{S,M_S,T, M_T, n} \nonumber \\
&=\frac{n(\Omega-n+\lambda+2)(\Omega+n-\lambda+4)}{4(n+2)} \nonumber \\
&\times\left[\frac{T(n+S+T+2)(n-S+T+1)}{2n(n+1)(2T+1)}\frac{(T-M_T)(T+M_T)}{T(2T-1)}+\frac{(T+1)(n+S-T+1)(n-S-T)}{2n(n+1)(2T+1)}\frac{(T+M_T+1)(T-M_T+1)}{(T+1)(2T+3)}\right] \nonumber \\
&+\frac{(n+4)(\Omega+n+\lambda+6)(\Omega-n-\lambda)}{4(n+2)}\nonumber \\
&\times\left[\frac{T(n-S-T+2)(n+S-T+3)}{2(n+3)(n+4)(2T+1)}\frac{(T-M_T)(T+M_T)}{T(2T-1)}+\frac{(T+1)(n-S+T+3)(n+S+T+4)}{2(n+3)(n+4)(2T+1)}\frac{(T+M_T+1)(T-M_T+1)}{(T+1)(2T+3)}\right] \nonumber
\eee
\bee
&\bra{S,M_S,T+2, M_T, n} A\dg_{0,1}(0,0)A_{0,1}(0,0)\ket{S,M_S,T M_T, n,} \nonumber \\
&=\sqrt{\frac{(T+M_T+1)(T-M_T+1)}{(2T+3)(T+1)}}\sqrt{\frac{(T+M_T+2)(T-M_T+2)}{(2T+3)(T+2)}}\nonumber \\
&\times \left[ \frac{n(\Omega-n+\lambda+2)(\Omega+n-\lambda+4)}{4(n+2)}\right. \sqrt{\frac{(T+2)(n+S+T+4)(n-S+T+3)}{2n(n+1)(2T+5)}}\sqrt{\frac{(T+1)(n+S-T+1)(n-S-T)}{2n(n+1)(2T+1)}}  \nonumber \\
&+\left. \frac{(n+4)(\Omega+n+\lambda+6)(\Omega-n-\lambda)}{4(n+2)}\sqrt{\frac{(T+2)(n-S-T)(n+S-T+1)}{2(n+3)(n+4)(2T+5)}}\sqrt{\frac{(T+1)(n-S+T+3)(n+S+T+4)}{2(n+3)(n+4)(2T+1)}} \right]  \nonumber
\eee
\bee
&\bra{S,M_S,T ,M_T, n+2} A\dg_{0,1}(0,0)A_{0,1}(0,0)\ket{S,M_S,T ,M_T, n} \nonumber \\
&=\sqrt{\frac{(n+2)(\Omega-n+\lambda)(\Omega+n-\lambda+6)}{4(n+4)}}\sqrt{\frac{(n+4)(\Omega+n+\lambda+6)(\Omega-n-\lambda)}{4(n+2)}} \nonumber \\
&\times \left[\sqrt{\frac{T(n+S+T+4)(n-S+T+3)}{2(n+2)(n+3)(2T+1)}\frac{T(n-S-T+2)(n+S-T+3)}{2(n+3)(n+4)(2T+1)}}\frac{(T-M_T)(T+M_T)}{T(2T-1)} \right.\nonumber \\
&+\left.\sqrt{\frac{(T+1)(n+S-T+3)(n-S-T+2)}{2(n+2)(n+3)(2T+1)}\frac{(T+1)(n-S+T+3)(n+S+T+4)}{2(n+3)(n+4)(2T+1)}}                                            \frac{(T+M_T+1)(T-M_T+1)}{(T+1)(2T+3)}\right]. \nonumber
\eee
\bee
&\bra{S,M_S,T+2, M_T, n+2} A\dg_{0,1}(0,0)A_{0,1}(0,0)\ket{S,M_S,T ,M_T, n} \nonumber \\
&=\sqrt{\frac{(n+2)(\Omega-n+\lambda)(\Omega+n-\lambda+6)}{4(n+4)}}\sqrt{\frac{(n+4)(\Omega+n+\lambda+6)(\Omega-n-\lambda)}{4(n+2)}}\sqrt{\frac{(T+2)(n+S+T+6)(n-S+T+5)}{2(n+2)(n+3)(2T+5)}} \nonumber \\
&\times \sqrt{\frac{(T+1)(n-S+T+3)(n+S+T+4)}{2(n+3)(n+4)(2T+1)}} \sqrt{\frac{(T+M_T+1)(T-M_T+1)}{(2T+3)(T+1)}}\sqrt{\frac{(T+M_T+2)(T-M_T+2)}{(2T+3)(T+2)}} \nonumber
\eee
\bee
&\bra{0,0,T-2 ,M_T-2, n-2}A\dg_{0,1}(0,-1)A_{0,1}(0,1)\ket{0,0,T, M_T, n} \nonumber \\
&=-\sqrt{\frac{(n+2)(\Omega+n+\lambda+4)(\Omega-n-\lambda+2)}{4n}}\sqrt{\frac{n(\Omega-n+\lambda+2)(\Omega+n-\lambda+4)}{4(n+2)}}\sqrt{\frac{(T-1)(n-1+T)(n+T)}{2(n+1)(n+2)(2T-3)}} \nonumber \\
&\times \sqrt{\frac{T(n+T+2)(n+T+1)}{2n(n+1)(2T+1)}} \sqrt{\frac{(T+M_T-2)(T+M_T-3)}{(2T-2)(2T-1)}}\sqrt{\frac{(T+M_T-1)(T+M_T)}{(2T-1)2T}}\nonumber
\eee
\bee
&\bra{0,0,T-2 ,M_T-2, n}A\dg_{0,1}(0,-1)A_{0,1}(0,1)\ket{0,0,T, M_T, n} \nonumber \\
&=-\sqrt{\frac{(T+M_T-2)(T+M_T-3)}{(2T-2)(2T-1)}}\sqrt{\frac{(T+M_T-1)(T+M_T)}{(2T-1)2T}}\nonumber \\
&\times \left[ \frac{n(\Omega-n+\lambda+2)(\Omega+n-\lambda+4)}{4(n+2)}\right. \sqrt{\frac{T(n+T+2)(n+T+1)}{2n(n+1)(2T+1)}}\sqrt{\frac{(T-1)(n-T+3)(n-T+2)}{2n(n+1)(2T-3)}}  \nonumber \\
&+\left. \frac{(n+4)(\Omega+n+\lambda+6)(\Omega-n-\lambda)}{4(n+2)}\sqrt{\frac{(T-1)(n+T+1)(n+T+2)}{2(n+3)(n+4)(2T-3)}}\sqrt{\frac{T(n-T+2)(n-T+3)}{2(n+3)(n+4)(2T+1)}} \right] \nonumber
\eee
\bee
&\bra{0,0,T ,M_T-2, n}A\dg_{0,1}(0,-1)A_{0,1}(0,1)\ket{0,0,T, M_T, n} \nonumber \\
&=\frac{n(\Omega-n+\lambda+2)(\Omega+n-\lambda+4)}{4(n+2)}\left[\frac{T(n+T+2)(n+T+1)}{2n(n+1)(2T+1)}\sqrt{\frac{(T-M_T+1)(T-M_T+2)}{2T(2T-1)}}\sqrt{\frac{(T+M_T-1)(T+M_T)}{2T(2T-1)}}\right.\nonumber \\
&+\left.\frac{(T+1)(n-T+1)(n-T)}{2n(n+1)(2T+1)}\sqrt{\frac{(T+M_T)(T+M_T-1)}{(2T+2)(2T+3)}}\sqrt{\frac{(T-M_T+1)(T-M_T+2)}{(2T+2)(2T+3)}}\right] \nonumber \\
&+\frac{(n+4)(\Omega+n+\lambda+6)(\Omega-n-\lambda)}{4(n+2)}\left[\frac{T(n-T+2)(n-T+3)}{2(n+3)(n+4)(2T+1)}\sqrt{\frac{(T-M_T+1)(T-M_T+2)}{2T(2T-1)}}\sqrt{\frac{(T+M_T-1)(T+M_T)}{2T(2T-1)}}\right.\nonumber \\
&+\left.\frac{(T+1)(n+T+3)(n+T+4)}{2(n+3)(n+4)(2T+1)}\sqrt{\frac{(T+M_T)(T+M_T-1)}{(2T+2)(2T+3)}}\sqrt{\frac{(T-M_T+1)(T-M_T+2)}{(2T+2)(2T+3)}}\right] .\nonumber
\eee

For $T_f=T-2$ the GT matrix element is
\bee
&&\sum_{\tilde{n}',M_S}\bra{0,0,T-2,T-2,\bar{n}'}\vec{\sigma}\tau^-\ket{1,M_S,T-1,T-1,\tilde{n}'}\cdot\bra{1,M_S,T-1,T-1,\tilde{n}'}\vec{\sigma}\tau^-\ket{0,0,T,T,n'}\nonumber\\
&\equiv&\sum_{n,\bar{n}} 2 c^{\bar{n}'}_{0,0,T-2,T-2,\bar{n}}c^{\tilde{n}'}_{1,0,T-1,T-1,\bar{n}}c^{\tilde{n}'}_{1,0,T-1,T-1,n}c^{n'}_{0,0,T,T,n}\nonumber \\&\times&\sqrt{\frac{(T-1)T(\bar{n}-T+2)(n-T+3)(n+T+1)(\bar{n}+T+2)}{(2T-1)(2T+1)}}.\nonumber
\eee

Within the perturbation theory the subject of interest is the GT matrix element as follows:
\bee
&&\sum_{M_S}\bra{0,0,4,2,4}\vec{\sigma}\tau^-\ket{1,M_S,3,3,4}\cdot\bra{1,M_S,3,3,4}\vec{\sigma}\tau^-\ket{0,0,4,4,4}
=-\frac{12}{\sqrt{7}}.\nonumber
\eee

\end{widetext}

%

%

\end{document}